\documentclass{arxiv}
\usepackage{amssymb}
\usepackage{pifont}
\usepackage{rotating}
\usepackage{epsfig}
\usepackage{subfigure}
\usepackage{rotating}
\usepackage{adjustbox}
\usepackage{graphicx}
\usepackage{epstopdf}
\usepackage{cite}
\usepackage{caption}
 \usepackage{color}
 \usepackage[normalem]{ulem}
 \usepackage{amsmath}

\title{\bf Forsaking your own: unveiling the delayed recognition of Garfield's work on the ``delayed recognition'' phenomenon}
\author{\small Tariq Ahmad MIR$^ {1,a,b}$ and Marcel AUSLOOS$^ {3,4,5,6,c,d,e}$\footnote{Deceased}\\}

\address{\footnotesize
$^1$Nuclear Research Laboratory, Astrophysical Sciences Division, Bhabha Atomic Research Centre, Srinagar-190 006, Jammu and Kashmir, India. \\$^a$ $e$-$mail$ $address$: taarik.mir@gmail.com\\\
$^b$ $e$-$mail$ $address$: tamir@barc.gov.in\\\
$^3$  GRAPES\footnote{Group of Researchers for Applications of Physics in Economy and Sociology} \\ rue de la Belle Jardini\`ere 483, B-4031, Angleur, Li\`ege, Belgium \\ 
$^c$$e$-$mail$ $address$:  marcel.ausloos@uliege.be \\
{$^4$ School of Business, University of Leicester,\\
Brookfield, 
Leicester, LE2 1RQ, UK\\   $^d$ $e$-$mail$ $address$: ma683@leicester.ac.uk
 \\$^5$   Universitatea Babeș-Bolyai, \\ Str. Mihail Kogălniceanu nr. 1, 400084, Cluj-Napoca,  Romania
  \\
  $^6$ Department of Statistics and Econometrics,  \\ Bucharest University of Economic Studies,  15-17 Dorobanti Avenue, \\ District 1, 010552, Bucharest, Romania, \\ $^e$  $e$-$mail$ $address$: marcel.ausloos@ase.ro}}
\begin{document}

\catchline{}{}{}{}{}

\maketitle


\section*{Abstract}
Delayed recognition (DR) implies that the full scholarly potential of certain scientific papers is recognized belatedly many years after their publication. Such papers are initially barely cited (sleep), and then suddenly, sometime in the future, their citation numbers burst (are awakened). After van Raan (2004a) 
called 
them ``Sleeping Beauties'' 
the 
DR phenomenon 
has drawn considerable attention. However, long before van Raan (2004a) 
Garfield 
studied the phenomenon in a series of articles  
from 1970 up to year 2004. In the present study we ask the pertinent question; Has the phenomenon of DR itself suffered the delayed recognition? In search of an answer we study the citation history of the Garfield (1980a) 
paper in which Garfield addressed DR directly for the first time. We find that the paper hardly received the 
attention befitting the Garfield's stature as an information scientist. 
Specifically, the paper received a meager of 10 citations up to the publication year of van Raan (2004a) and was then, in 2007, feebly awakened from its deep sleep of twenty-eight years 
receiving 20 citations in next four years; up to 2010. Being the undisputed giant of information science that even Garfield's paper on DR can suffer DR is hardly anticipated. 

\section*{Keywords}

Eugene Garfield; Delayed recognition; Sleeping beauty



\section{Introduction}
Certain scientific 
papers, amongst the vast number of them published, 
are initially looked down. 
The issue has been much delved upon 
by sociologists of science. Barber (1961) provides examples of 
new scientific ideas resisted by scientists themselves on account of being against the science in vogue, professional specialization and standing of the discoverers in the social system of science. 
Merton (1963) discusses the resistance to the systematic study of multiples discoveries, again by scientists themselves, highlighting how the scientist of higher professional standing prevails when there are priority disputes; who was the first to make the discovery under contest, whereas lesser known ones get pushed over. 
Cole (1970) 
found that while it is the quality of the paper that determines its reception the reputation of the author, based on his previous work, does affect  its acceptance. Stent (1972), discussing the 
discovery of hereditary nature of DNA, attributes the lack of its appreciation, and of scientific discoveries in general, to 
its prematurity in that it can't be logically incorporated into the larger cannon of knowledge reasons for which, in this specific case, were pointed out to be purely technical by Wyatt (1975). 

Those studies by sociologists, e.g., Barber (1961) and Merton (1970) are qualitative, except somewhat for Cole (1970), whereas these by scientists, Stent (1972) and Wyatt (1975), are related to the specific examples of the ignored papers from their own specialization. The quantitative analysis of the broader phenomenon 
was first taken up by Garfield aided by his unique position of being the owner of the {\it``Science Citation Index''} (SCI) the only citation database available then. 
Garfield (1970), a two page note in Current Contents (CC) magazine, 
debunked the alleged lack of acknowledgement of the importance of Mendel's seminal work on genetics. Notice that  the latter was then purported to be a classic case 
of a work, having gone unnoticed; the true worth was realized after a delay of more than four decades. 

Drawing on the work of sociologists and scientists,
we alluded to above, 
Garfield (1980a, 1980b) revisited the phenomenon, 
a decade after Garfield (1970), 
using the term DR 
for the first time\footnote{Garfield (1970) used the term ``ignore'' for failure to foresee the importance of such papers. Garfield (1970) and Cole (1970), published their view in the same year, but do not cite each other despite both using SCI in their analysis.}. Garfield (1980b, 1980c) justified the raison d'être of SCI,  against criticism 
exuding confidence that its availability 
will aid not only in identifying useful work but would also prevent their unintentional neglect and duplication. 
Furthermore,  Garfield (1989a) through quantitative analysis of  citation histories contained in SCI unraveled the scientific works which had been ignored initially for a considerable period of time for reasons that sociologists of science had much discussed. 
Interestingly, Garfield actively engaged with the readers of the CC calling upon them to report any such works, thereafter publishing their letters in the magazine, and then analyzing SCI data to confirm that the reported works 
have indeed suffered DR (Garfield, 1989c, 1989d).  Garfield (1992) used the citation data of their work to predict the future Nobel prize winners  and confirmed the DR, or not, 
of the 
winning work of those who have already been awarded the prize (Garfield, 1985a, 1986b). 

After vigorously pursuing it 
for over two decades Garfield grew skeptical of DR phenomenon  
(Gl\"{a}nzel \& Garfield, 2004). Nevertheless, the subject bounced back to prominence in year 2004 itself after van Raan (2004a) rechristened papers having suffered DR as "Sleeping Beauties" (SBs).  By analogy with the stories by Perrault (1697) and by the Grimm brothers (1812) such a paper whose citation triggers the resurgence of a SB has its ``prince". The citation accumulation  of SB papers is peculiar in that they are initially only occasionally cited, or not cited at all, for a long time (sleeping time) after their publication and  some time in the future, following a citation by the "prince", their citations start increasing rapidly (awakening time). This citation trajectory is in stark contrast to that of a typical scientific paper whose citations peak with in the first few years of its publication (Gl\"{a}nzel \& Schoepflin, 1995; Gl\"{a}nzel et al., 2003; Costas et al., 2010) and then continuously decline over the years and finally stop accumulating.

Following van Raan (2004a) 
the study of SBs, or DR phenomenon as Garfield called it, has attracted much attention. 
Scientometric studies 
among others include unearthing of SBs and of their respective princes 
through scanning of large citation databases (Gl\"{a}nzel et al., 2003; van Raan, 2004a; Braun et al., 2010; Ke et al., 2015), developing different criteria and quantitative methods for their unearthing (Winnink et al, 2019; Miura et al., 2021), identifying different types of SBs (Li \& Ye, 2012; van Raan, 2015), unraveling the reasons of their DR and also for their awakening (van Dalen \& Henkens, 2005; Wang et al., 2012; Li \& Shi, 2016;  Ye \& Bornmann, 2017), strategies for preventing DR of important work in future (Jiang et al., 2024), using data other than citation; altmetrics, for their identification (Hou \& Zhang, 2023) and pointing out the extreme cases of those SBs which the analysis of large databases fail to capture (Van Calster, 2012; Mir \& Ausloos, 2018). Sometimes, the discussion pertains to very specific fields (Fazelli-Varzaneh et al., 2021).

Unlike van Raan (2004a) Garfield's work on DR hardly elicited any interest from the wider scientific community, - a response unanticipated given his acclaimed stature as an information scientist. 
Intriguingly 
the lone paper, amongst his work on DR, featuring in the list of his 50 most cited papers is Gl\"{a}nzel \& Garfield (2004) 
when they expressed skepticism of the phenomenon (Panday \& Gupta, 2021). Despite Garfield's pursuance of more than two decades, DR 
is not counted amongst 
inspirations that influenced his scientific work (Bornmann et al., 2018). 
Furthermore, neither are any of his articles on the phenomenon included amongst his 10 
most under-cited yet influential articles (Rousseau \& Hu, 2018) 
nor are they amongst his most read publications (Thelwall, 2017). 
On the other hand, Chen (2018) found that citations to Garfield (1980a) have bursted upto year 1985; - which however we reveal is purely due to self-citations, six of them, in this period. 

Therefore, in the present study, we ask the pertinent question: Has the Garfield's work on DR been a victim of DR itself?, i.e.,  an undeserving fate which he tried to bring to the attention of a wide scientific community. 
Furthermore, we wonder 
if the experts in the field, including Garfield himself, have forgotten their own backyard while looking for the dormitory of SBs in other areas of knowledge. In search for an answer we study Garfield's entire body of work on DR phenomenon 
published over the years. 
Specifically we analyze the citation history of Garfield (1980a) paper in which Garfield used the term DR for the first time to describe the publications whose true potential is belatedly recognized. 

In order to do so, first we test the citations received by this  1980a paper against Garfield's own criteria (Garfield 1989a, 1990a) for identifying a paper as a case of DR. Next, we apply the criteria  recently prescribed by van Raan (2004a) to check if Garfield (1980a) paper is a SB. We also calculate the beauty coefficient and awakening time for this SB as per the procedure of Ke et al. (2015). 

Surprisingly, we reveal that Garfield (1980a) paper has struggled for its entire citation life before achieving the 10-fold of the citations accumulated in the first decade of its publication. Furthermore, the maximum number of citations for any given year was never crossing the count of one for an incredible 23 years up to year 2002. Thus, we find that the paper was in a state of deep sleep for a period of 28 years up to 2007. Furthermore, without citing Garfield (1980a) van Raan (2004a) is a prince who had just a glimpse of its SB, thereby leading only to its ''feeble awakening''.  


\section{Data}
Obtaining citations data for Garfield's work is a challenge given that most of it  
appeared in CC magazine, - his self-owned and self-edited publication 
(Bornmann et al., 2018). Bereft of peer review process, the benchmark of scholarly literature and basis for its indexing, 
the metadata corresponding to Garfield's work are not correctly reported by Web of Science (WoS) and Scopus, i.e., the two main bibliometric databases (Jasco, 2018). 
Therefore, neither Scopus 
nor WoS can be trusted, the latter ironically, for the oeuvre of its own founder (Hu \& Rousseau, 2017). Indeed Jasco (2010) found that WoS does not provide correct information on the bibliography and citations received for Garfield's contributions in the CC. Pertinently, Jasco (2010) revealed how WoS wrongly attributed all the citations received by the 52 articles in 1980 issue of CC to a single paper, which was an editorial about the new building of the Institute for Scientific Information. Notice that the issue contains the paper Garfield (1980a) the number of citations to which is the subject matter of present investigation. 


The limitations of Google Scholar (GS) (Jasco, 2005) and the problematic coverage of Garfield's oeuvre by the Scopus and WoS implies that none of the three citation databases can be used in isolation. Therefore, to build a comprehensive dataset of papers citing Garfield (1980a) we combine the data from the three databases. Each database was searched for the title of Garfield (1980a) paper on September 15, 2025. The overall citation count to the paper from Scopus, WoS and GS respectively turned out to be 45, 93 and 205. The smaller number of yearly citations 
enabled us to inspect each of the search results 
so that we 
keep at bay any duplicate or 
incorrect citations. The later concern is particularly true for GS data which we corrected for one duplicate self-citation, Garfield \& Welljams-Dorof (1992), for year 1992 and for two wrongly attributed self-citations, Garfield (1990a), for 1998. 
Next to ensure that none of the citing papers is counted multiple times as a citation for any given year we compared the yearly search results of the three databases. 
For example, for 2007 there respectively are one, two and three citations to Garfield (1980a) in WoS, GS and Scopus. However, one citing paper, Gl\"{a}nzel (2007), is listed in, is common to, all the three datasets so we count it only once whereas another one, Simkin \& Roychowdhury (2007), is found in GS and Scopus but not in WoS and therefore it is again counted once only. On the other hand, Levitt \& Thelwall (2007) is listed in Scopus only. So for year 2007 though there are a total of six citations listed in the three databases only three of them are unique citations. In this way for every year we identified unique citations to Garfield (1980a). We found fifteen such citations from Scopus and seven from WoS not listed in GS. 
From the three databases we build a dataset of 214 unique papers citing Garfield (1980a). 

Knowing that 
only robust data can lead to reliable conclusions we have to ensure that the incomplete coverage of Garfield's oeuvre by three databases 
does not pollute our analysis. 
Admittedly when it comes to looking for 
information on 
Garfield nothing can be more authentic than that 
provided by Garfield himself. The Garfield Library of the University of Pennsylvania  ({\color{blue}{https://garfield.library.upenn.edu/}}) is an exhaustive source of information on the biographical material, academic achievements and investments of Garfield. To crosscheck our search results from GS, WoS and Scopus, we turned to this eponymous library. The database has customized search box powered by Google 
on its home page. 
A search dated September 26, 2025, for the title of Garfield (1980a), as we had done on the three databases, 
displayed 286 results. 


We downloaded the PDF's of search results and inspected them all, one by one, to find out which ones actually cite the Garfield (1980a) paper. 
It was observed that in many of his publications Garfield just alluded 
to whereas in a few he discussed the DR phenomenon, - without explicitly citing Garfield (1980a). We consider, as per the standard scientometric practice, that the latter paper was cited only when reference is included in the bibliography of the citing paper. In all we found 26 papers of Garfield 
focused on the DR phenomenon. 
Among these 26 articles, 23 appeared in CC; 2 papers, i.e., Garfield (2000) and Gl\"{a}nzel \& Garfield (2004), appeared in The Scientist (TS). Like CC, the latter magazine  was also founded (in 1986) and edited by Garfield. 
Garfield \& Welljams-Dorof (1992) is his only article on DR that appeared in a journal not founded, owned or edited by him. 

The articles published in CC magazine were later compiled into 15 volumes of the Essays of an Information Scientist (EIS) ({\color{blue}{https://garfield.library.upenn.edu/essays.html}}). 
During our investigation,  we found that some confusion persists for the bibliographic information when the same articles are referred to from the two sources and many times are misquoted. Therefore, in Table 1 we list Garfield's  all publications on the DR phenomenon providing complete information of papers published in CC and their corresponding information in EIS. 
Furthermore, we also show which of his 24 subsequent articles on DR phenomenon cite Garfield (1980a). As can be seen from Table 1, there are 
20 self-citations. 
 To complete the data 
we add these self-citations to the citations collected from GS, WoS and Scopus, for the respective years of publication. This dataset of 234 citing papers, 214 from three citation databases plus 20 self-citations from CC, forms the basis of our analysis. The time series is displayed on Fig.1. 

Of the Garfield's 26 articles on DR phenomenon none of his 23 articles from CC are listed as having cited Garfield (1980a) in WoS and Scopus whereas Garfield (1990a), the lone article listed in GS, is wrongly listed as a citation for year 1998. Gl\"{a}nzel \& Garfield (2004), article from ``The Scientist'' magazine is his only article covered by all the three databases whereas Garfield \& Welljams-Dorof (1992) is covered by GS only. These findings align well with prior studies on the inadequate coverage of Garfield's oeuvre by the three databases (Jasco 2010, 2018).

\begin{table}[h]
 \tbl{Garfield's papers on DR arranged in chronological order, Note: CC stands for Current Contents, EIS for Essays of an Information Scientist, TS for The Scientist and TMS for Theoretical Medicine and Bioethics}
{\begin{tabular}{@{}ccccccccccc@{}} \toprule
\hphantom{00}No. & Paper & CC & CC &  CC & ESI  & ESI & Cites & 
\\
\hphantom{00} & & Issue & pages& Date & Volume & pages & Garfield (1980a)  
&\\
\botrule
1  &Garfield (1970) \hphantom{00} & 2 & 5 - 6 & January 14 & 1 & 69 - 70 &  
 \\\\
2 &  Garfield (1980a) \hphantom{00} & 21 & 5 - 10 & May 26 & 4 & 488 - 493 &  
 \\ \\ 
3 & Garfield (1980b) \hphantom{00} & 44 & 5 - 10 & November 3 & 4 & 660 - 665 & Yes 
 \\ \\
4 & Garfield (1980c) \hphantom{00} & 38 & 5 - 13 & September 22 & 4 & 609 - 617 & Yes 
 \\ \\
5 & Garfield (1983) \hphantom{00} & 45 & 5 - 14 & November 7 & 6 & 363 - 372 & Yes 
 \\ \\
6 &  Garfield (1984a) \hphantom{00} & 45 & 3 - 15 & November 5 & 7 & 353 - 365 & Yes 
 \\ \\
7 & Garfield (1984b) \hphantom{00} & 50 & 3 - 17 & December 10 & 7 & 405 - 419 & Yes 
 \\ \\
8 & Garfield (1985a) \hphantom{00} & 7 & 3 - 10 & February 18 & 8 & 60 - 67 & No 
 \\ \\
9 & Garfield (1985b) \hphantom{00} & 47 & 3 - 18 & November 25 & 8 & 444 - 459 & Yes 
 \\\\
10 & Garfield (1986a) \hphantom{00} & 23 & 3 - 8 & June 9 & 9 & 182 - 187 &  Yes 
 \\\\
11 & Garfield (1986b) \hphantom{00} & 44 & 3 - 12 & November 3 & 9 & 336 - 345 &  Yes 
 \\\\
12 & Garfield (1988) \hphantom{00} & 11 & 3 - 12 & March 14 & 11 & 75 - 84 &  Yes 
 \\\\
13 & Garfield (1989a) \hphantom{00} & 23 & 3 - 9 & June 5 & 12 & 154 - 159 & Yes 
 \\\\
14 & Garfield (1989b) \hphantom{00} & 3 & 3 - 10 & January 16 & 12 & 16 - 17 & Yes 
 \\\\
15 &  Garfield (1989c) \hphantom{00} & 38 & 3 - 8 &  September 18 & 12 & 264 - 269 & Yes 
 \\\\
16 &  Garfield (1989d) \hphantom{00} & 49 & 3 - 8 & December 4 & 12 & 345 - 350 & No 
  \\\\
17 & Garfield (1990a) \hphantom{00} & 9 & 3 - 9 & February 26 & 13 & 68 - 74 & Yes 
  \\\\
18 & Garfield (1990b) \hphantom{00} & 8 & 3 - 13 & February 19 & 13 & 57 - 67 & Yes 
  \\\\
19 &  Garfield (1990c) \hphantom{00} & 5 & 3 - 5 & January 29 & 13 & 29 - 31 & Yes 
 \\\\
20 &  Garfield (1990d) \hphantom{00} & 43 & 3 - 14 & October 22 & 13 & 387 - 398 & Yes 
 \\\\
21 & Garfield (1990e) \hphantom{00} & 35 & 3 - 13 & August 27 & 13 & 316 - 326 & No 
 \\\\
22 &  Garfield (1992) \hphantom{00} & 35 & 3 - 12 & August 31 & 15 & 127 - 136 & Yes 
 \\\\
23 & Garfield \& \hphantom{00} &  &  & & &  & Yes 
 \\
 & Welljams-Dorof (1992) \hphantom{00} & TMB  &  & & &  & 
 \\\\ 
24 & Garfield (1993) \hphantom{00} & 5 & 3 - 12 & February 1 & 15 & 228 - 237 & Yes 
 \\\\
25 &  Garfield (2000) \hphantom{00} & TS & & & & & No 
 \\\\
26 & Gl\"{a}nzel \& Garfield (2004) \hphantom{00} & TS &  & & &  & Yes 
 \\
\botrule
\end{tabular} \label{ta3}}
\end{table}

\section{Analysis and results}
We trace out the citation trajectory of Garfield (1980a) in Fig. 1. In the inset plot, we show the yearly citations received by the paper, including the self-citations identified in Table 1, whereas the corresponding yearly citation counts without self-citations are shown in the main plot. Recall that depiction of the citation frequencies in two plots is intentional for the application of two different criteria for identifying a paper as being a case of DR (i) one set by Garfield himself and (ii) the recent one prescribed by van Raan (2004a). 

Garfield set quantitative criteria for identification of DR  papers in terms of the temporal accumulation of their citations, in 1989, a decade after Garfield (1980a). According to Garfield (1989a) for a highly cited paper to qualify 
as an example of DR it must satisfy two criteria (i) 
low citation frequencies for the first 5 or more years, - with more than 10 years preferable 
(ii) low initial citation frequency 
of about 
an average of one cite per year for a typical paper. Later Garfield (1990a) set more specific criteria: (i) a threshold of 10 or fewer citations to a paper at age 10 and (ii) 
a 10-fold increase in citations at age 20. Importantly, Garfield 
does not exclude self-citations to an article from the analysis, as is the current practice, 
when deciding on its DR. 

From the inset plot in Fig.1, it can be seen that Garfield (1980a) 
received 21 citations up to year 1990. This count amounts to an average of about $1.91(\frac{21}{11})$ 
citations per year,- a number higher than average of one citation per year as required by Garfield (1989a). Note that this citation count 
includes 16 self-citations two out of which, Garfield (1980b, 1980c), 
came 
in the year of publication itself. Therefore, the paper does not satisfy the first criterion; 10 or fewer citation in first decade of publication.   Having received only 4 citations, 3 of them being self-citations (2 for year 1992 and 1 for 1993), the next decade [1991-2000] of its citation life was even more miserable. Obviously the citation count of 4 is a far cry from 210 citations, 10 fold of 21, 
those, accumulated in the first life time decade, as demanded by Garfield's second criterion. 

The paper attracted 31 citations in third decade, 2000-2010, including one self-citation, Gl\"{a}nzel \& Garfield (2004). Although this citation count is larger than 25, accumulated in the first two decades, nevertheless it is still very short of 210, the 10-fold increase in citations. 
With a total citation count of 234 
the Garfield (1980a) paper has struggled in the entire course of its citation life, to meet the second criterion set, Garfield (1990a), by its own author.

\begin{figure}
\hspace*{-25pt}
\subfigure{\label{}\includegraphics[width=.6\linewidth, height=1.05\linewidth,  angle=90,]{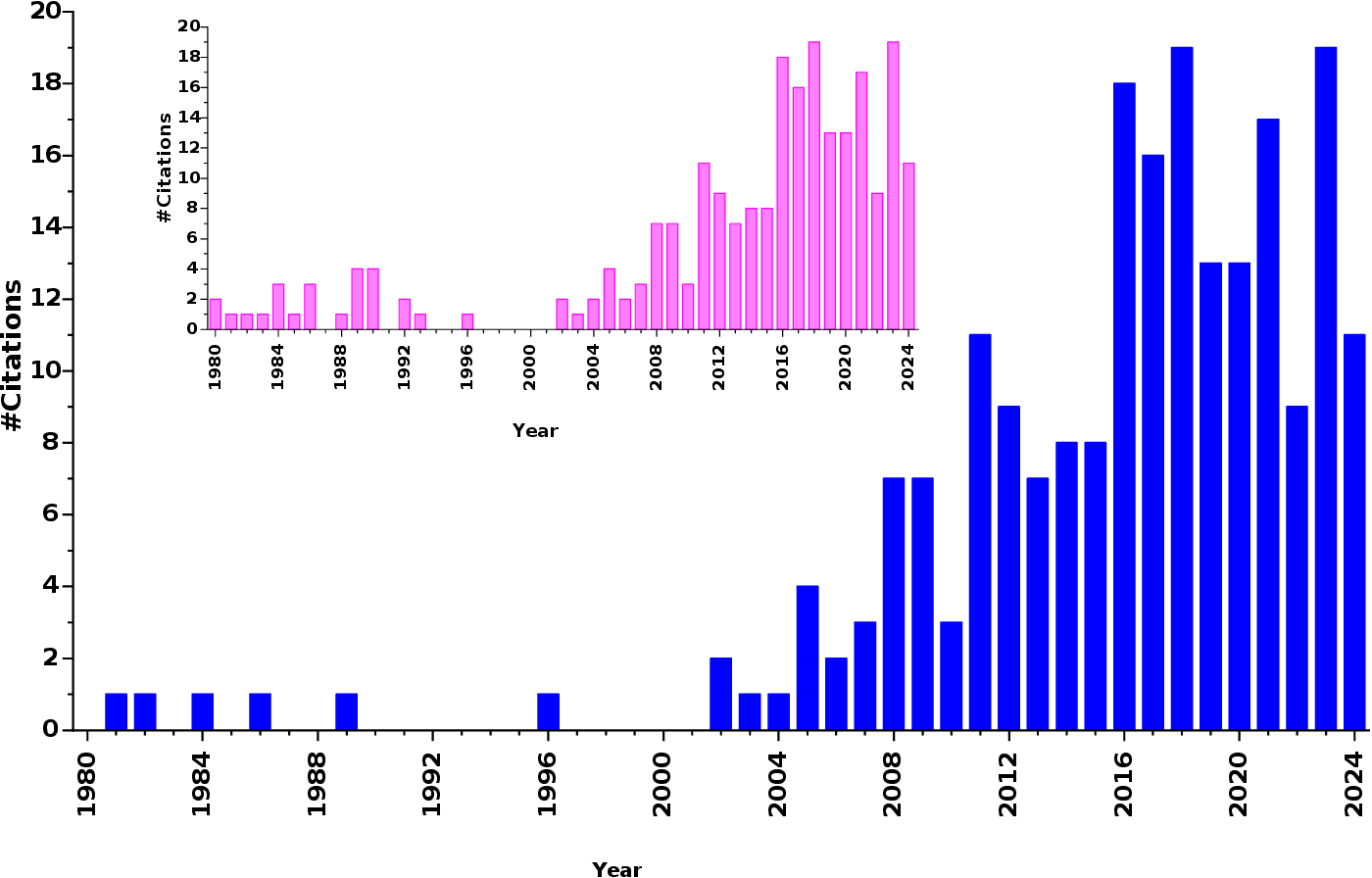}}
\vspace*{-10pt}
\caption{Yearly citations to Garfield (1980a) (main plot excludes self citations). Inset plot is for all citations including self-citations.}
\end{figure}

But for 
his self-citations 
the DR phenomenon was forgotten for two and a half decades after Garfield (1980a). The phenomenon came back live after it was revisited by 
van Raan (2004a), albeit using a different terminology,  
calling papers having suffered DR as ``Sleeping Beauties''. 
For their definition/identification van Raan (2004a) prescribed a criteria different than the one set in Garfield (1989a, 1990a). The van Raan criteria are based on three main variables: (1) depth of sleep, i.e., the article receives at most one citation on average per year (deep sleep), or between one to two citations on  average per year during a specific period (less deep sleep); (2) the length of sleep, i.e., the duration of the above period; and (3) the awakening intensity, i.e., the number of citations per year, during four years following the sleeping period. Later on van Raan (2015) prescribed that (4) SBs on average be cited more than five times a year in the awakening stage. Note that van Raan (2004a) criterion differs from that of Garfield (1980a) in excluding the self-citations from the analysis.
Next, we use the these criteria 
to asses whether Garfield (1980) paper is a SB or not. 

From the citation trajectory shown in the main plot of Fig. 1, it can be seen that Garfield (1980a) received only 10 citations in 25 years, 1980-2004, of its publication and therefore with an average of less than one, $0.4(\frac{10}{25})$, citation: it was in a state of deep sleep for this period. 
It is incredible that up to year 2002, for 23 years, the maximum number of yearly citations never crossed the citation count of one. 

In respect to the awakening intensity, the paper awoke in year 2007, 28 years after its publication, when it received 20 citations; with yearly citations of 3, 7, 7, 3, up to 2010. Thus as per van Raan (2004a) criteria, Garfield (1980a) paper on DR phenomenon is a clear case of DR itself and hence a SB. 

The comparison of the two plots in Fig. 1 clearly indicate that for 14 years, 1980-1993, Garfield pursued the phenomenon vigorously; 19 self citations, but the wider scientific community, - only 5 citations in this period, failed to appreciate his idea of DR. Notice that due to complete absence of self-citations the two plots are exactly same for 1994-2003 period. The lull in self-citations is indicative of a decade of hiatus by Garfield in this research field again possibly fueled by lack of any appreciation. This conjecture is supported by the skepticism of DR phenomenon 
expressed in Gl\"{a}nzel \& Garfield (2004).
From year 2005 onwards both the plots in Fig. 1 have the same number of yearly citations. This is due to the absence of any further self-citations to Garfield (1980a) after  Gl\"{a}nzel \& Garfield (2004). 
Interestingly, a year after in 2005 the citations to the paper increased to 4, maximum count for any given year upto this year, for the first time since its publication. Then onwards the paper started to attract more than one citations regularly reaching a maximum of 19 in year 2018 an indication of renewed interest in DR phenomenon. 

Notice that Gl\"{a}nzel et al., (2003) and Gl\"{a}nzel \& Garfield (2004) analyzed about half a million papers of the 1980 edition of the SCI 
and concluded that nearly all significant research is well cited within the first three to five years of publication. Both 1980, the year of edition of their data source, and 2004, the year of publication of their analysis are ironic. In the former year, Garfield (1980a) discussed the phenomenon in the historical perspective, coining the term DR, whereas in the latter year when Garfield called the phenomenon only a myth, following van Raan (2004a), he saw its rebirth. Furthermore, had Garfield included Garfield (1980a) paper in his analysis, 
he would have been surprised to know that up to 2004 other researchers had given his paper on DR, a dismal of 10 citations, and that it had got a new lease of citation life after the year 2004, he would have reached an opposite conclusion on the occurrence of the phenomenon.

\section{Discussion}

The modest increase in citations to Garfield (1980a) after year 2004 follows the publication of van Raan (2004a) who borrowed the term ``Sleeping Beauty`` from French folklore to describe the phenomenon and literally took the field of scientometrics by storm. Sugimoto \& Mostafa (2018) question such a terminology for being culturally insensitive. van Raan's criteria for the identification of SB 
are broadly similar to those of Garfield's for DR;  both prescribe certain period of no citations for a paper (sleeping period) followed by sudden increase in its citations (awakening). The fact that van Raan (2004a) received immediate attention whereas response to Garfield's work has been largely lukewarm 
clearly indicates that 
use of 
the vivid names did work in favor 
of the formers reception 
as is evident from the exponential growth of literature on the subject. For further quantitative depiction of the latter fact, we collected, on September 15, 2025, 
the yearly citations received by van Raan (2004a) from GS; for a quick visual appreciation, we show these in Fig. 2. As can be seen the paper received prompt attention, i.e., citations without even a delay of a single year, after its appearance; therefore it was an instant hit. In fact, the paper received 14 citations; with one self-citation van Raan (2005), in very second year of its publication.

\begin{figure}
\hspace*{-25pt}
\subfigure{\label{}\includegraphics[width=1.05\linewidth, height=.6\linewidth,  angle=0,]{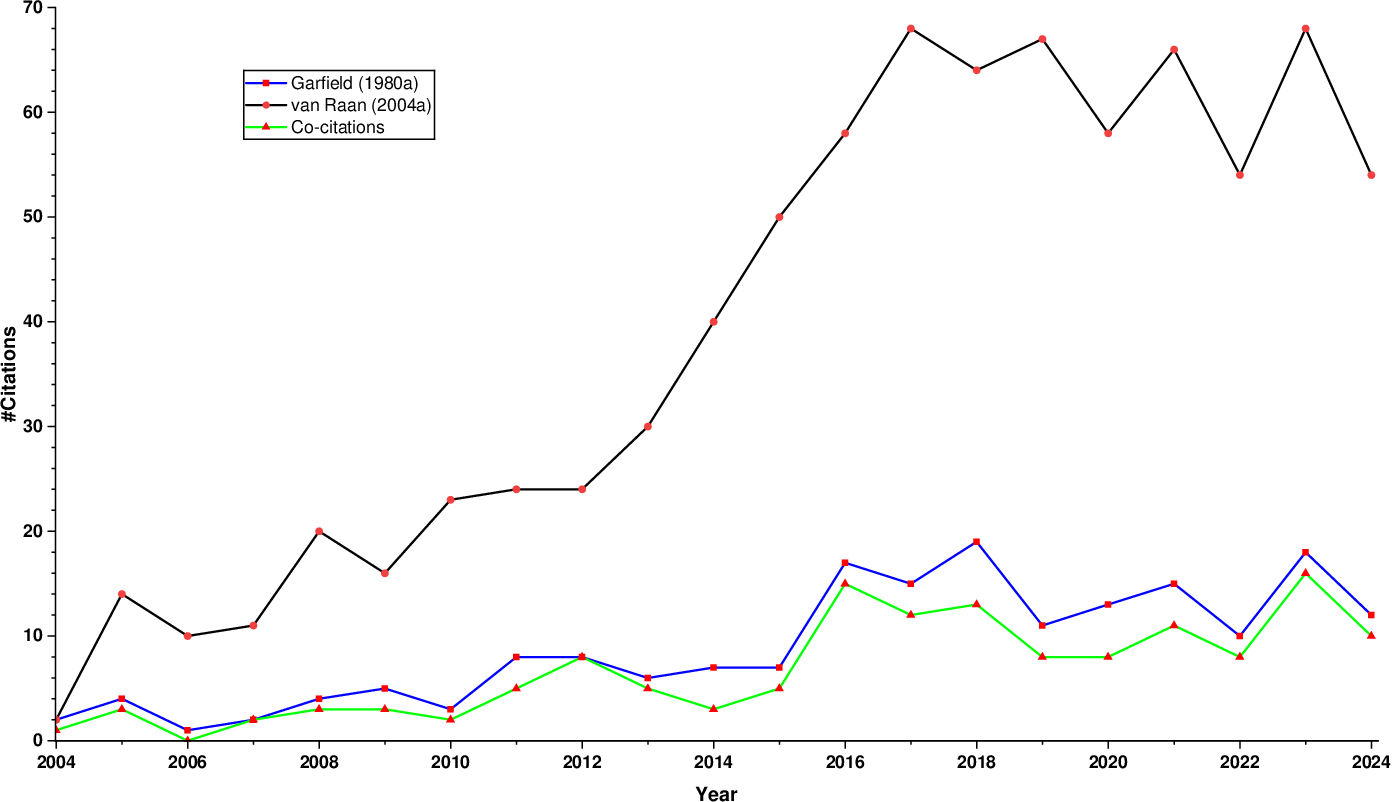}}
\vspace*{-10pt}
\caption{Comparison of yearly citations to Garfield (1980a) and van Raan (2004a) [self-citations are included for both papers].}
\end{figure}

We show the yearly citations, blue curve in Fig. 2, received by Garfield (1980a) after  
van Raan (2004a). 
The citation revival of Garfield (1980a) after year 2004 is obvious. 
Nevertheless, a casual comparison of the their citation curves reveals that the yearly citation counts of Garfield (1980a) are no match to that of van Raan (2004a), - with a trend continuing for their entire citation journey. It had taken 9 self-citations for the former paper 
to accumulate a citation count of 13 over a period of 9 years upto 1988. The comparison with the delay of not a single year for van Raan (2004a) to accumulate, 14 citations, more than that count does not demand any other quantitative measure for  reaching conviction. Furthermore, the wait for Garfield (1980a) to achieve this citation goal gets longer by an incredible two and half decades; 25 years up to 2005, when its self-citations are discounted.
Garfield (1980a) during its 45 years of citation life has accumulated only 234 citations (all databases) more than the corresponding 262 van Raan (2004a) has collected upto 2015 i.e. in only eleven years of its publication.

Recall that van Raan (2004a) does not cite Garfield (1980a); the former paper has a very short bibliography, - with three references only, but was cited in the same year (van Raan, 2004b), when reviewing the quantitative measures for the "evaluation of science". 
{Lack of a citation and much smaller number of yearly citations to Garfield (1980a) compared to van Raan (2004a), blue and black curves in Fig. 2, on the surface, suggest that later paper is not the ''prince`` who awakened the former. 
Such a conjecture is negated when the co-citations of the two are taken into account. Through out their citation journey the curve for the co-citations of the two papers, green color in Fig. 2, is seen to be closer to the citation curve of Garfield (1980a)  indicating a high rate of co-citations;- peaks for years 2012, 2016 and 2023 are readily notable. Garfield (1980a) received a total of 187 citations after van Raan (2004a) out of which the two papers are co-cited 141 times  i.e.- about 75\% $(\frac{141}{187})$, a high co-citation rate indeed. Therefore, van Raan (2004a) is a ''prince`` who just had a glimpse, without a kiss, on  the Garfield (1980a), - the SB. Furthermore, even after 2007, the year of its  awakening, the yearly citations of the latter are very modest compared to those of its ''prince`` therefore indicating that the SB has had only a ''feeble awakening``. 

Having established that van Raan (2004a) is its prince we now calculate the beauty coefficient of Garfield (1980a). Introduced by Ke et al., (2015), the index takes into account the peak citations and the year of its occurrence in the citation curve of a SB. Given that the citation curve of Garfield (1980a), main plot in Fig. 1, has two peaks of 19 citations for year 2018 and 2023, there respectively are two beauty coefficients viz. 159.55 and 144.62 (Li \& Ye, 2016). Nevertheless, both the values of the beauty coefficient yield the same awakening time of 26 years corresponding to year 2006. The awakening year of 2006 
is in close agreement to 2007, obtained from van Raan (2004a)'s criteria. Pertinently Ke et al., (2015) showed that Garfield has another much severe SB to his credit i.e. Garfield (1955), which paper had slept for 50 years before being awakened around year 2000.

In conclusion, although highly influential, Garfield (1980a) is another example of these highly under-cited articles yet it is not covered in  Rousseau \& Hu (2018). Perhaps Garfield (1991b) had an inkling of his un(under)-citedness when he consoled a scientist that there is no shame in being so. 
\section{Conclusion}
Garfield is an undisputed giant of information science having multiple contributions to the field. Previously, befitting to his stature, the impact of his contributions on a wider field of ''science of science`` has been widely discussed. Here we studied the response to "delayed recognition", a phenomenon Garfield investigated for over more than two decades. In order to do so, we perform a thorough analysis of citation history of the Garfield (1980a) paper which coined the term ''delayed recognition``. Ironically we found it to have suffered the same fate which it had started to bring to the notice of our wide scientific community. Despite frequently self-citing his paper Garfield could not prevent 28 years delay of its recognition through citations. Having been in a state of deep sleep for 28 years the paper is clearly a "Sleeping Beauty" seeing the revival of its citations following van Raan (2004a) kiss. These results are surprising 
when seen in the light of  the theme of the Garfield (1980a) paper and his work on the phenomenon. 



\end{document}